\documentclass[aps,prl,twocolumn,superscriptaddress]{revtex4-1} 
\usepackage{graphicx}
\usepackage{subfigure}
\usepackage{bm}
\usepackage{amsmath,amsthm,amssymb}
\usepackage{amsfonts}    
\usepackage{color}      
\usepackage{subfigure}  
\usepackage{epsfig}
\usepackage{float}
\usepackage{epstopdf}
\usepackage{appendix}
\usepackage{multirow}

\usepackage[colorlinks=true, pdfborder=001, linkcolor=blue, anchorcolor=blue, citecolor=blue, urlcolor=blue]{hyperref}

\begin{document}
\title{Stable regular solution of Einstein-Yang-Mills equation}
\author{Yuewen Chen}

\email{yuewen\_chern@amss.ac.cn}
\affiliation{Yau Mathematical Sciences Center, Tsinghua University, Beijing 100084, China}

\author{Shing-Tung Yau}
\email{Corresponding author: yau@math.harvard.edu}
\affiliation{Yau Mathematical Sciences Center, Tsinghua University, Beijing 100084, China}
\affiliation{Yanqi Lake Beijing Institute of Mathematical Sciences and Applications, Beijing 101408, China}
\affiliation{ Department of Mathematics, Harvard University, Cambridge, MA 02138, USA }

\date{\today }

\begin{abstract}
 In this letter, we  find the first dynamically  stable non-singular solution of spherically symmetric $SU(2)$ Einstein-Yang-Mills  equation. This solution is regular at $r=0$ and asymptotically flat.  
 Since the Yang-Mills field strength decay exponentially, the Einstein-Yang-Mills particle perhaps can be used as a candidate for dark matter.
  \end{abstract}

\maketitle

\section{Introduction}
 
 In 1988, a global nontrivial static nonsingular particle-like solution of coupled Einstein-Yang-Mills (EYM) equation was found by  Bartnik and McKinnon numerically \cite{Bartnik}. It aroused a great of interest in GR community. It was established rigorously by Smoller-Wasserman-Yau {\it et.al}  \cite{s2,s1,s5}. It was also found there is an infinite  black solutions different coupling constants. Since it violated the familiar no hair theorem, much excitement was generated. However, it was soon demonstrated by Strauumen-Zhou \cite{un1}  that these solutions are not dynamically stable and therefore not physical. In this letter, we find the first dynamically stable non singular solution of Einstein-Yang-Mills. Since the field strength of Yang Mills of our solution decay exponentially,  we believe such particle solution can be  a potential candidate for dark matter.
 
 After  Bartnik-McKinnon's pioneering work, a large number of soliton and black hole solutions of spherically symmetric EYM equations were found \cite{2, 3, Kun}. The critical behavior of spherically symmetric collapse of EYM equations were studied in \cite{cho1,cho2,cho3,cho4}. 
 Smoller and Yau {\it et.al}  proved rigorously  that the $SU(2)$ EYM equations admit an infinite family of black-hole solutions with a regular event horizon \cite{s4}. However, the colored black hole found numerically by Bizon is also  unstable\cite{un2,un3}. The nonlinear stability was studied in \cite{un4} and the work provided numerical evidence for the instability of the colored black hole solutions.

 The new idea in this letter, is  to study   the evolution of the time dependent EYM equation rather than only studying  the static equation \cite{Bartnik}. Our main new result is the observation that for suitable initial data, the  EYM equations would have a  non trivial steady state solution.  Furthermore, we show this solution is stable under  linear perturbation.
We write the spherically symmetric metric as 
$$ds^2=-A e^{2Q} dt^2+\frac{1}{A} dr^2+r^2 d\Omega^2.$$

The Yang-Mills curvature tensor is given by
\begin{align*}
F=&W_t \tau_1 dt \wedge d\theta+W_t \tau_2  dt\wedge \sin \theta d \phi+W' \tau_1 dr \wedge d\theta\\
     &+W' \tau_2  dr\wedge \sin \theta d \phi -(1-W^2) \tau_3 d \theta \wedge \sin \theta d \phi,
\end{align*}
where $\tau_i$ are the Pauli matrices.
The evolution version of Einstein-Yang-Mills equation are given by
\begin{align}
(\frac{1}{Ae^{Q}} W_t)_t &=(A e^Q W_r)_r\nonumber\\
&\ \ \ \ \ +\frac{e^Q}{r^2}W(1-W^2),\label{eq1} \\
A_t &=-\frac{4}{r}W_t W_r A,\label{eq2}\\
rA_r+A(1+2(W_r^2+\frac{1}{A^2e^{2Q}} W_t^2)) &=1-\frac{1}{r^2}(1-W^2)^2,\label{eq3}\\
rQ_r &=2(W_r^2+\frac{1}{A^2e^{2Q}} W_t^2) \label{eq4}.
\end{align}

 
 We need a new coordinate transformation $$x=\ln (r).$$
 We define the auxiliary functions $\Box, S, \alpha, \theta, q$ and $c$, respectively,  as  follows
 \begin{align}
 \Box&=2(W_r^2+\frac{1}{A^2e^{2Q}} W_t^2),\\
 S&=1-\frac{1}{r^2}(1-W^2)^2,\\
 \alpha&=e^Q,\\
 \theta_x &=-\Box, \\
 q_x &=1+\Box,\\
 c&=A \alpha.
 \end{align}
 
 To evolution  Einstein-Yang-Mills system, we choose equations (\ref{eq1}), (\ref{eq3}) and (\ref{eq4}).
 Then, we rewrite the EYM system as 
 
 \begin{align}
 c (\frac{1}{c } W_t)_t &=\frac{c }{r}(\frac{c }{r} W_x)_x+\frac{c \alpha}{r^2}W(1-W^2) ,\label{eq:1}\\
 (Ae^q)_x &=S e^q,\label{eq:2} \\
 (\alpha e^\theta)_x &=0.\label{eq:3}
 \end{align}

\section{Stability }

\begin{figure*}[htbp]
\centering
\subfigure[]{
\begin{minipage}[h]{0.5\linewidth}
\includegraphics[width=1.1\linewidth]{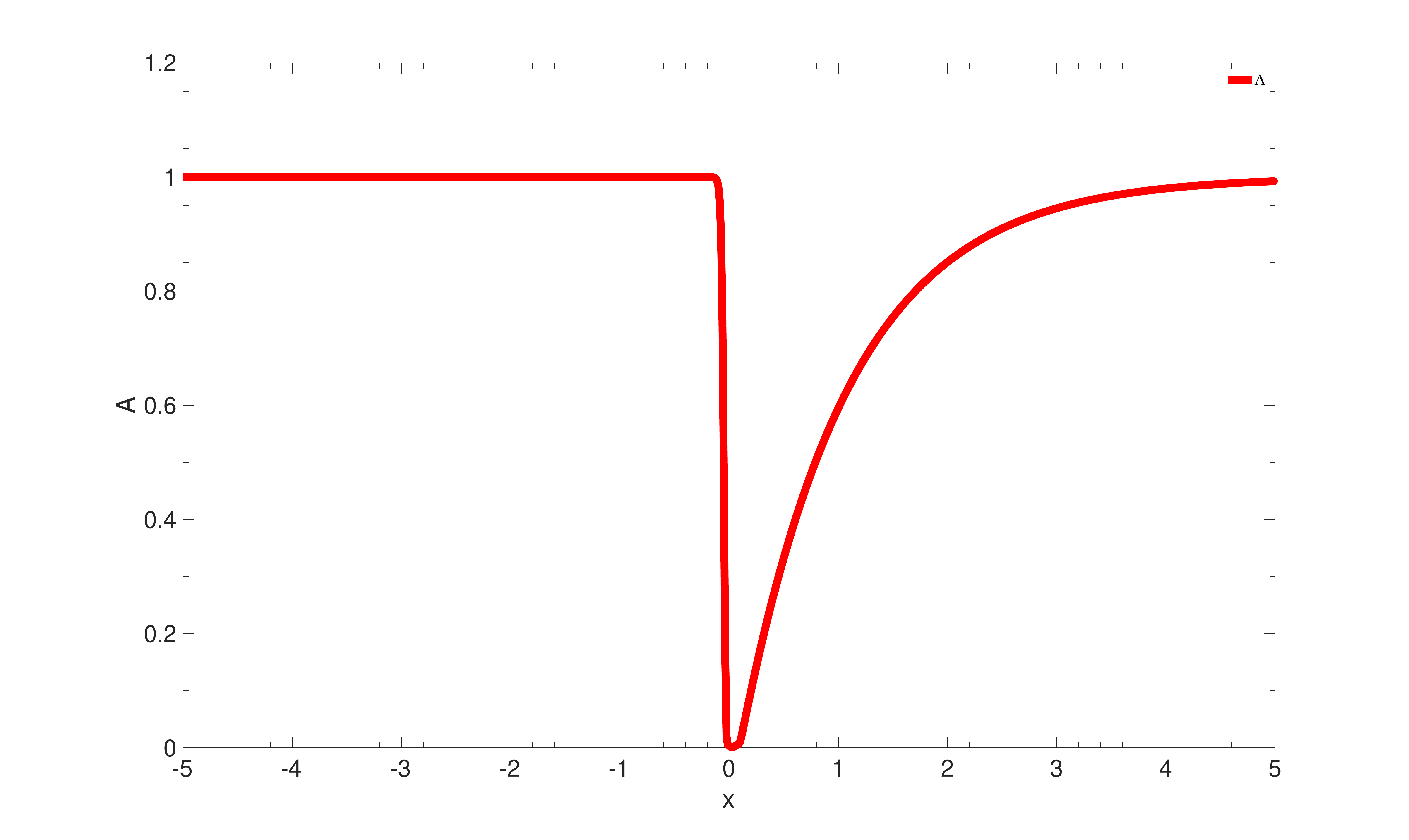}
\end{minipage}}
\subfigure[]{
\begin{minipage}[h]{0.48\linewidth}
\includegraphics[width=1.1\linewidth]{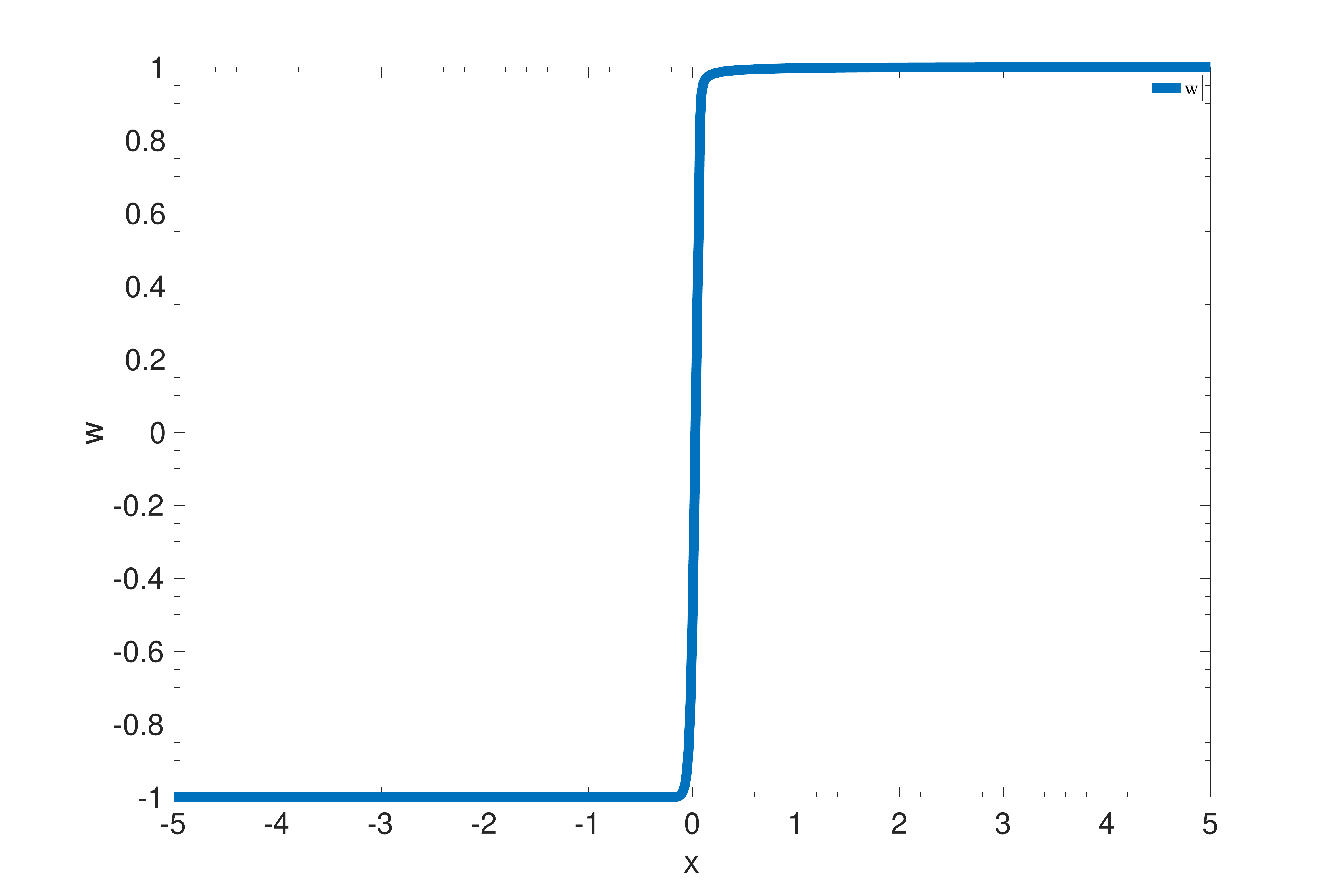}
\end{minipage}}
\subfigure[]{
\begin{minipage}[h]{0.48\linewidth}
\includegraphics[width=1.1\linewidth]{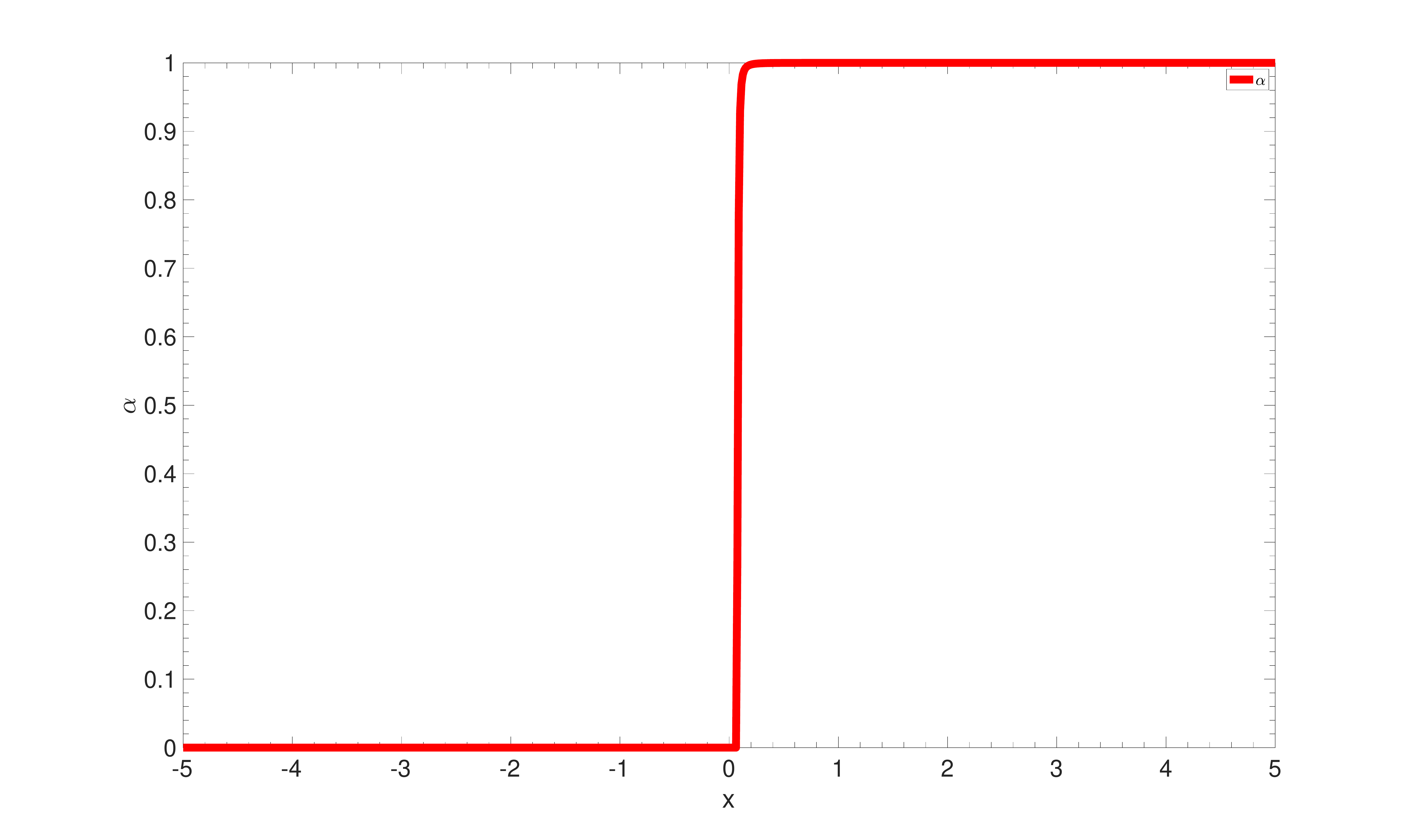}
\end{minipage}}
\subfigure[]{
\begin{minipage}[h]{0.46\linewidth}
\includegraphics[width=1.12\linewidth]{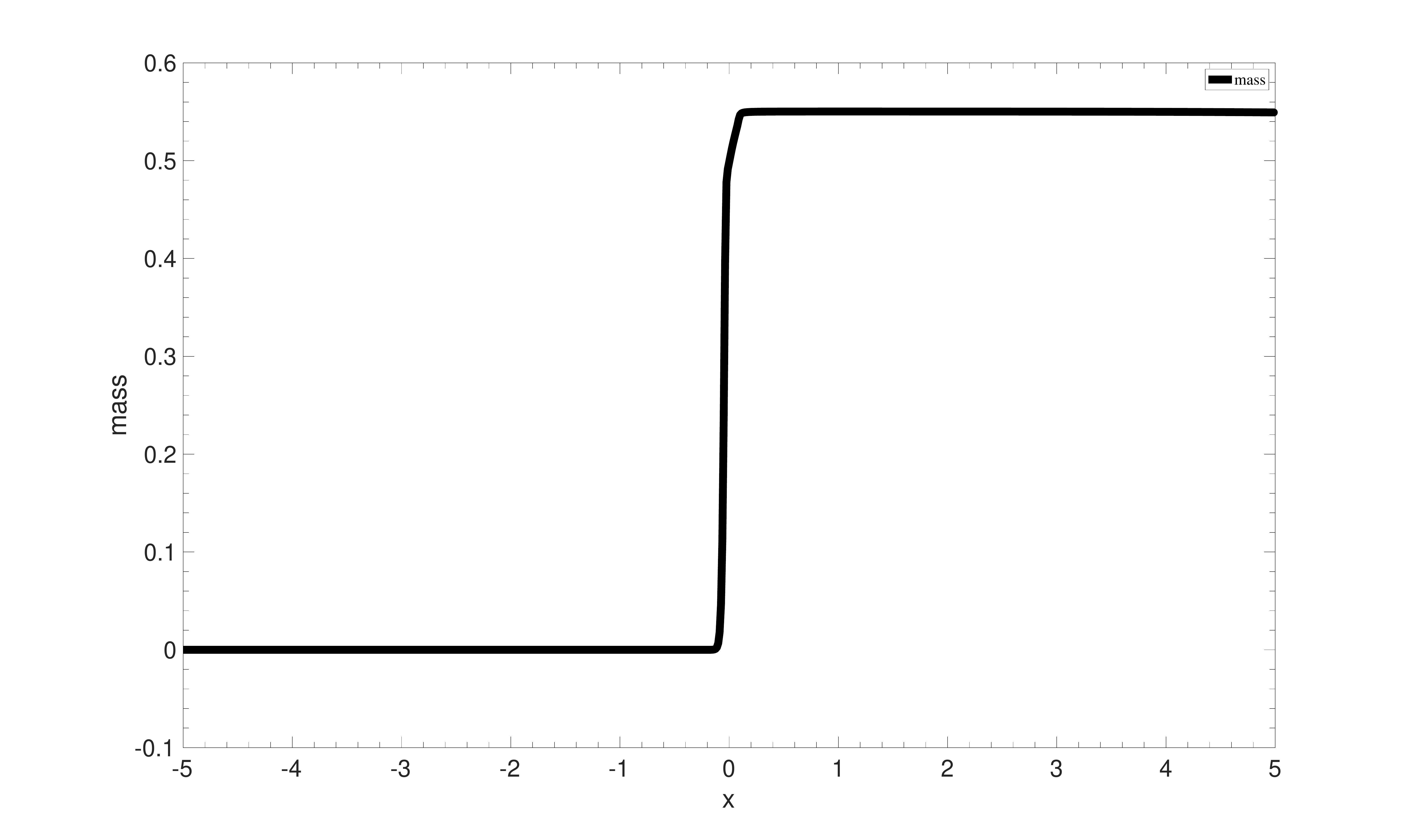}
\end{minipage}}
\caption{ (a): The $A$ is regular $r=0$. There is no horizon and $\min(A)=0.0004$. Using different initial data, one can get different smooth steady state solution, such as $W(x,0)=\tanh(10(x-0.23))$, the $\min(A)=0.0026.$ (b): $W$ is smooth and approximates $1$ in the far field and close to -1  near $r=0.$
(c): $\alpha$ closes to 0 as $x \to -\infty$ ($r \to 0$).  (d): The mass increases to the total mass $M=0.54.$
} \label{fig1}
\end{figure*}

We consider the  following initial  data for $W$
$$W(x,0)=\tanh(10(x-0.03)), \quad W_t(x,0)=0.$$
We show the steady state solution of the EYM equations in Fig. \ref{fig1}.

We consider the linear perturbation of EYM  systems in this section.   For convenience, we  write the  metric as 
$$ ds^2=-A \alpha^2 dt^2 +\frac{1}{A} dr^2 +r^2(d\theta^2 +\sin ^2\theta d\phi^2)$$
and denote the $SU(2)$ gauge potential by 
\begin{align*} {\frak A}&= a_0 \tau_r dt+ a_1 \tau_r dr +(1+W) (\tau_\theta \sin \theta d \phi -\tau_\phi d \theta)\\
&\ \ \ \ \ +u(\tau_\theta d \theta +\tau_\phi \sin \theta d \phi),
\end{align*} 
where $A,\alpha, a_0, a_1,W, u$ are  functions of $r, t$ and
\begin{align}
\tau_r &=\tau_1 \sin\theta  \cos \phi +\tau_2 \sin \theta \sin \phi +\tau_3 \cos \theta , \\
\tau_\theta &=\tau_1 \cos \theta \cos \phi +\tau_2 \cos \theta \sin \phi -\tau_3 \sin \theta , \\
\tau_\phi &=-\tau_1 \sin \phi +\tau_2 \cos \phi
\end{align} 
with $\tau_i, i=1, 2, 3$ are the usual Pauli spin matrices. For more details, see Ref. \cite{Ge3}.
In the static case, we set $a_0=a_1=u=0.$
We will consider the spherically symmstric perturbations for our regular solution obtained above. In history, many people study the instability  of Bantnik-McKinnon solutions and  colored black holes solutions.   The unstable mode found bu Zhou-Straumann\cite{un2}  belong to perturbations within the original Bantnik-McKinnon ansatz. Beside these even-parity modes there  is a second  class of exponentially growing modes\cite{Ge1,Ge2}, which was called "sphaleron-like" instability. 
The complete set of perturbation can be decoupled into two groups: even sector (``gravitational") and odd sector (``sphaleron"), because  the parity transformation $\theta \to \pi -\theta, \phi \to \phi +\pi$ is a symmetry operation \cite{Ge1,Ge2}.

 The  even-parity sector perturbation ansatz are 
\begin{align}
W(t,r)&=W_0+\varepsilon \tilde{W}(r) e^{i\sigma t},\\
A(t,r)&=A_0+\varepsilon \tilde{A}(r)e^{i\sigma t},\\
\alpha(t,r) &=\alpha_0 + \varepsilon \tilde{\alpha}(r) e^{i\sigma t},
\end{align}
where $W_0,A_0,\alpha_0$ are the  background solution  obtained above (See Fig. 1).
 
 We obtain the  following ODE eigenvalue problem 
$$-\sigma^2  \tilde{W} =A_0^2 \alpha_0^2 \tilde{W}_{rr}+\alpha_0^2 C_3 \tilde{W}_r+\tilde{W} \alpha_0^2 (C_1-\frac{4}{r}C_2A_{0}W_{0,r}),$$
where
\begin{align*}
C_1&=A_0(\frac{1-3W_0^2}{r^2}-\frac{4W_0(W_0^2-1)}{r^3}W_{0,r}), \\
C_2&=2A_0W_{0,rr}+\frac{1}{r^2} W_0(1-W_0^2),\\
C_3&=\frac{A_0}{r}(1-A_0-\frac{(1-W_0^2)^2}{r^2}).
\end{align*}

The  first eigenvalue is given by $$\sigma^2=24607.8079188817,$$
which  show that the $\sigma \in \mathbb{R}$, and show the solution of the linear equation would not grow.  

Next, we consider the odd-parity   sector perturbation   
\begin{align}
a_0(t,r) &=\varepsilon \tilde{a}_0(r) e^{i \sigma t}, \\
a_1(t,r) &=\varepsilon \tilde{a}_1 (r)e^{i \sigma t} ,\\
u(t,r)&=\varepsilon \tilde{u} (r)e^{i \sigma t}.
\end{align}

The odd-parity (``sphaleron") perturbation equations are given by
\begin{align}
\alpha_0^2 A_0^2(- \tilde{u}_{rr} +W_0 \tilde{a}_{1,r} -\frac{W_0^2-1}{A_0 r^2} \tilde{u}) &=\sigma^2 \tilde{u}, \\
\frac{2}{r^2}\alpha_0^2 A_0 (W_{0,r} \tilde{u} -W_0 \tilde{u}_{r} +W_0^2\tilde{a}_1) &=\sigma^2 \tilde{a}_1.
\end{align}
The  first eigenvalue of the  above   system  is $$\sigma^2=24359.6486332623.$$
Hence, the $\sigma \in \mathbb{R}$ and  sphaleron sector will not grow.

\section{Acknowledgments}
We would like to thank Professor   George Lavrelashvili  for valuable remarks and comments. 

\section{Appendix: Light cone and Curvature}
Here, we will  study the geometry structure  of the smooth solution. First, we consider the  null geodesic, which are given by 
\begin{align}
\frac{dt}{dr}=\pm \frac{1}{A\alpha}
\end{align}
or, in $x$ coordinate 
$$t_x=\pm \frac{e^x}{A\alpha}.$$
We integral the null geodesic and show the light cones in Fig. 2.

\begin{figure}[H]
		\centering
	\includegraphics[width=9.6cm]{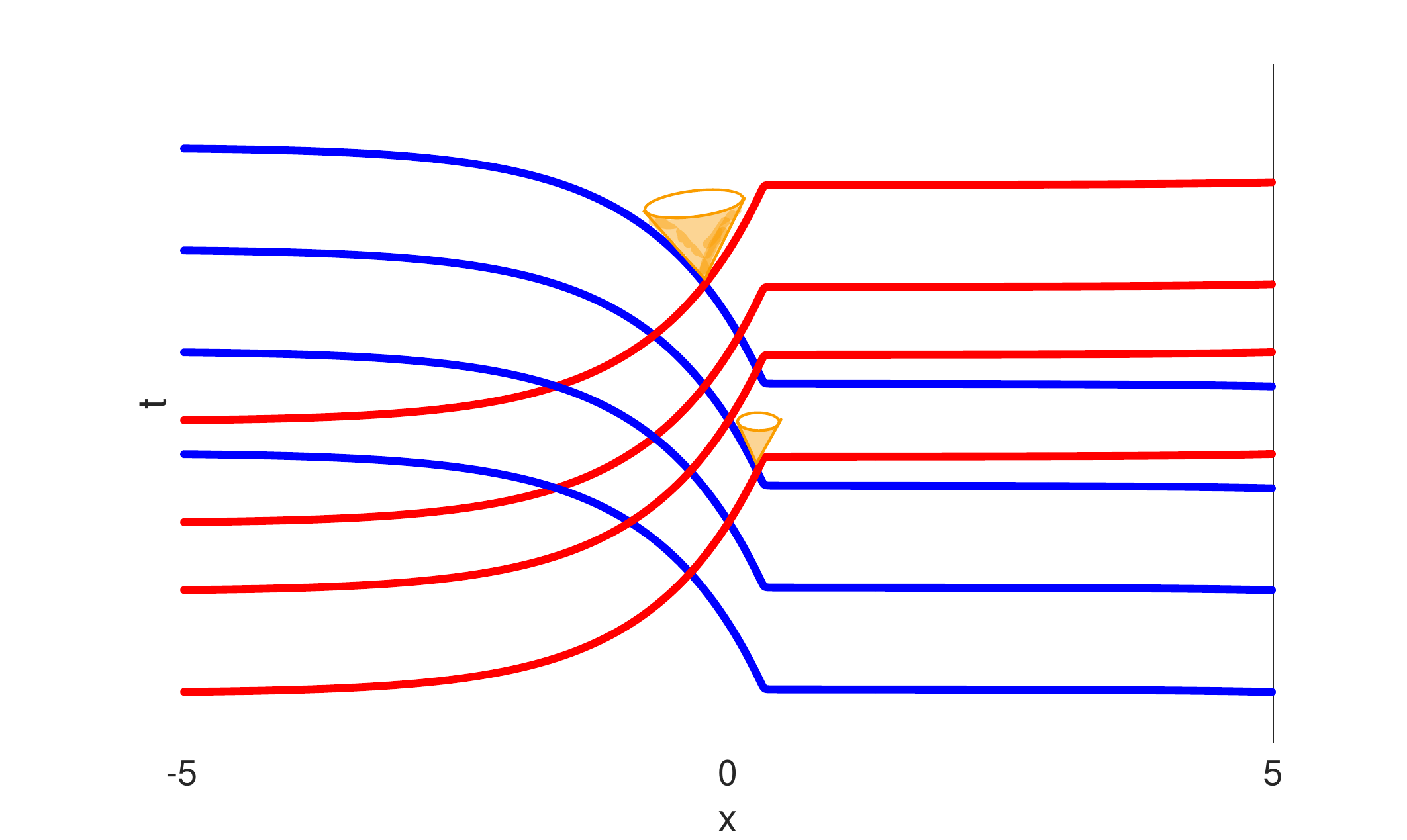}\\
	\caption{Light cones of the smooth solution.}
	\label{fig3}
\end{figure}

The Ricci  scalar curvature is computed in  Fig. \ref{fig3},  and we found it is $0$ everywhere.
\begin{figure}[H]
		\centering
	\includegraphics[width=9.6cm]{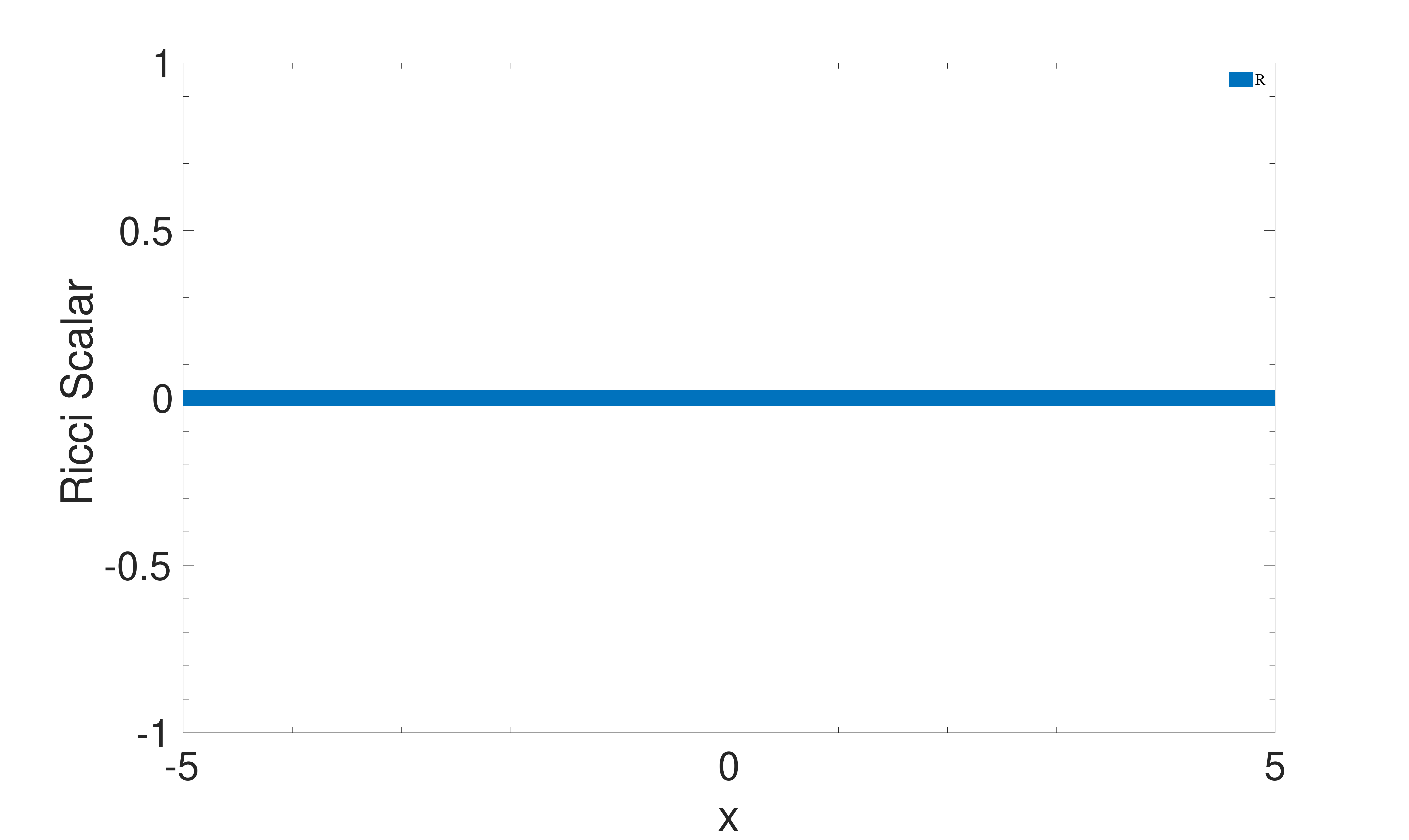}\\
	\caption{The Ricci  scalar curvature is 0 everywhere.}
	\label{fig3}
\end{figure}

\end{document}